\documentclass[aps,amssymb,showpacs,twocolumn]{revtex4-1}
\usepackage{graphicx}
\usepackage{lipsum}
\usepackage{hyperref}

\begin{document}

\title{X-ray absorption spectroscopy and magnetic circular dichroism studies of L1\(_0\)-Mn-Ga thin films}

\author{M.\ Glas}
\email{mglas@physik.uni-bielefeld.de}
\homepage{www.spinelectronics.de}
\author{C.\ Sterwerf}
\author{J.\ M.\ Schmalhorst}
\author{G.\ Reiss}
\affiliation{Thin Films and Physics of Nanostructures, Bielefeld University, Germany}
\author{D.\ Ebke}
\affiliation{Thin Films and Physics of Nanostructures, Bielefeld University, Germany}
\affiliation{Max Planck Institute for Chemical Physics of Solids, Dresden, Germany}
\author{C.\ Jenkins}
\author{E.\ Arenholz}
\affiliation{ALS Berkeley, CA, USA}

\date{\today}

\begin{abstract}
Tetragonally distorted \(\rm{Mn}_{3-x}\rm{Ga}_x\) thin films with \(0.1< x < 2\) show a strong perpendicular magnetic anisotropy and low magnetization and thus have the potential to serve as electrodes in spin transfer torque magnetic random access memory. Because a direct capping of these films with MgO is problematic due to oxide formation, we examined the influence of a CoFeB interlayer, and of two different deposition methods for the MgO barrier on the formation of interfacial MnO for \(\rm{Mn}_{62}\rm{Ga}_{38}\) by element specific X-ray absorption spectroscopy (XAS) and magnetic circular dichroism (XMCD). A highly textured L1\(_0\) crystal structure of the Mn-Ga films was verified by X-ray diffraction (XRD) measurements. For samples with e-beam evaporated MgO barrier no evidence for MnO was found, whereas in samples with magnetron sputtered MgO MnO was detected, even for the thickest interlayer thickness.  Both XAS and XMCD measurements showed an increasing interfacial MnO amount with decreasing CoFeB interlayer thickness. Additional element specific full hysteresis loops determined an out-of-plane magnetization axis for the Mn and Co, respectively. 
\end{abstract}


\maketitle

\section{Introduction}
The continuous miniaturization of spintronic devices requires consistently new materials providing high spin polarization and Curie temperature to achieve high tunnel magneto resistance (TMR) effects at room temperature and low magnetic damping, low saturation magnetization to lower the critical current for spin-transfer torque (STT) switching.\cite{Graf:2011jj} In addition, a large perpendicular magnetic anisotropy is required to guarantee reliable data retention. Examples of commonly used thin films with perpendicular magnetic anisotropy are \(\mathrm{(Co/Pd)_n}\)\cite{Kugler:2011dz} or \(\mathrm{(Co/Pt)_n}\)\cite{Park:2008cc} multilayers, which take advantage of the large magnetic anisotropy induced by interfaces. Another approach to induce a perpendicular magnetic anisotropy is the utilization of ferromagnetic materials in combination with rare earth elements. The integration of such films into perpendicular magnetic tunnel junctions (pMTJs) or spin valves (SPVs) was reported by Nakayama et al.\ \cite{Nakayama:2008bu} and Li et al.\ \cite{Li:2011by} for \(\rm{Tb}\)-\(\rm{CoFe}\) and \(\rm{Tb}\)-\(\rm{Co_2FeAl}\). Both the multilayer structures and the ferromagnetic materials in combination with rare earth elements provide a high anisotropy energy density \(\mathrm{K = M_S H_K/2}\), with \(\mathrm{H_K = H_{Sat} + 4 \pi M_S}\).\cite{Carcia:1985gp,Johnson:1999jh,Lee:2010ii} \(\mathrm{H_{Sat}}\) is defined as the saturation field in hard magnetic axis direction. But due to the high spin-orbit interaction of noble metals and also rare earth elements, the damping constant \(\mathrm{ \alpha }\) is large (\(\alpha_{\rm{Co/Pd}} \approx 0.09\))\cite{Kato:2012tb} and thus a high critical current density for STT switching can be expected. Combined with the challenging deposition of multilayers and the high reactivity of rare earth elements this may hinder the wide spread use of these materials in application. To overcome these obstacles new materials have to be developed. A promising material is the \(\mathrm{Mn_{3-x}Ga}\) compound, which is predicted to fulfill \cite{Graf:2011jj} the criteria for STT switching and stability. In the \(\mathrm{X_2YZ}\) Heusler-type \(\mathrm{D0_3}\) phase, the X and Y positions are occupied with \(\mathrm{Mn}\) atoms and the \(\mathrm{Ga}\) atoms occupy the Z positions. Wurmehl et al.\ predicted \(\mathrm{Mn}_3\mathrm{Ga}\) to be a half-metallic, fully compensated ferrimagnet.\cite{Wurmehl:2006db} The tetragonally-distorted L1\(_0\) and \(\mathrm{D0_{22}}\) crystallographic phases provide an easy magnetization axis perpendicular to the film plane if the distortion axis is in the growth direction. The L1\(_0\) phase can be found for \( x > 1.2\). However, between \(0.15 \le x \le 1.06\) the D0\(_{22}\) crystal structure can be achieved.\cite{NIIDA:1996uh}
Sakuma et al.\ reported a theoretical magnetic Mn moment of 2.51\(\mu_B\) for MnGa in the L1\(_0\) structure.\cite{Sakuma:1998hd} Hasegawa and Tsuboya experimentally obtained a magnetic moment of 1.7\(\mu_B\). This decrease in the magnetic moment is based on a \(\rm{Mn}\) content higher than one. Thereby Mn atoms occupy Ga positions and couple antiferromagnetically to the initial Mn atoms. In addition, an anisotropy constant of \(K_{u} = 2.6 \times 10^{7}\,\rm{erg/cm}^3\) was measured.\cite{Sakuma:1998hd}\newline
Furthermore, Mizukami et al.\ reported a very low magnetic damping constant of 0.008 for \(\mathrm{{Mn_{1.54}Ga}}\).\cite{Mizukami:2011ft}\newline
Kubota et al.\ observed a TMR effect of 6\,\% (18\,\%) at 300\,K (10\,K) for L1\(_0\) ordered \(\mathrm{Mn}_{54}\mathrm{Ga}_{46}\) based MTJs.\cite{Kubota:2011dz} Additionally Bai et al.\ reported a first principle study of the magnetic and transport properties of \(\mathrm{Mn_{3-x}Ga}\)/\(\mathrm{MgO}\)/\(\mathrm{Mn_{3-x}Ga}\) magnetic tunnel junctions.\cite{Bai:2012iz} Bai pointed out, that only for low Mn concentration (\(\rm{Mn_{2}Ga}\)) a high TMR effect should be observable. For higher Mn concentrations (\(\mathrm{Mn}_{3}\mathrm{Ga}\)) spin filtering effects vanish. The reported TMR effects are low compared with other Heusler materials.\cite{Ebke:2010ci} An important reason of these lower TMR effects could lie in the low spin polarization of the s-electrons of Mn\(_{3-x}\)Ga, which are mostly responsible for incoherent tunneling.
Additionally the oxidation of the Mn could cause the low TMR and/or the lattice mismatch between the Mn-Ga bottom layer and the MgO barrier. In this paper we investigate the oxidation of the Mn during the deposition of the MgO barrier and the influence of an additional CoFeB interlayer to improve the material systems for applications.

\section{Experimental}
The samples were prepared by conventional DC and RF magnetron sputtering. The \(\mathrm{Mn_{1.63}Ga}\) thin films were deposited in a UHV system from a \(\mathrm{Mn_{50}Ga_{50}}\) sputtering target. The \(\mathrm{Ar}\) pressure was \(\mathrm{1.3 \times 10^{-3}\,mbar}\) and the growth rate was set to approximately 0.4\,nm/sec. To achieve a tetragonally distorted lattice, the deposition temperature was 520\,\(^{\circ}\)C. The Mn-Ga layers were deposited on \(\mathrm{SrTiO_3}\) (001) substrates. The cubic \(\mathrm{SrTiO_3}\) exhibits a lattice parameter of 3.905\,\AA. Due to the low lattice mismatch (\(<1\,\%\))  highly textured thin films are obtained.\cite{Glas:2013vc} An additional interlayer of \(\mathrm{Co_{40}Fe_{40}B_{20}}\) (CoFeB) was deposited at room temperature on the \(\mathrm{Mn_{1.63}Ga}\) thin film, with thicknesses from 0 to 2\,nm. Finally, the layers were capped by a 2\,nm thick \(\mathrm{MgO}\) layer. Two different deposition procedures were used for the \(\mathrm{MgO}\) tunnel barrier, magnetron sputtering and e-beam evaporation.\newline
X-ray absorption spectroscopy (XAS) and X-ray magnetic circular dichroism (XMCD) measurements were performed at beamline (BL) 6.3.1 of the Advanced Light Source in Berkeley, USA. The \(\rm{Mn}\) \(L\) edges were investigated by surface sensitive total electron yield in normal incidence. XMCD spectra were obtained by applying a magnetic field of 1.5\,T along the X-ray beam direction using elliptically polarized radiation with degrees of polarization of +66\,\%. The sample temperature was 300\,K. \newline

\section{Results}
\begin{figure}[b!]%
\includegraphics[width=\linewidth]{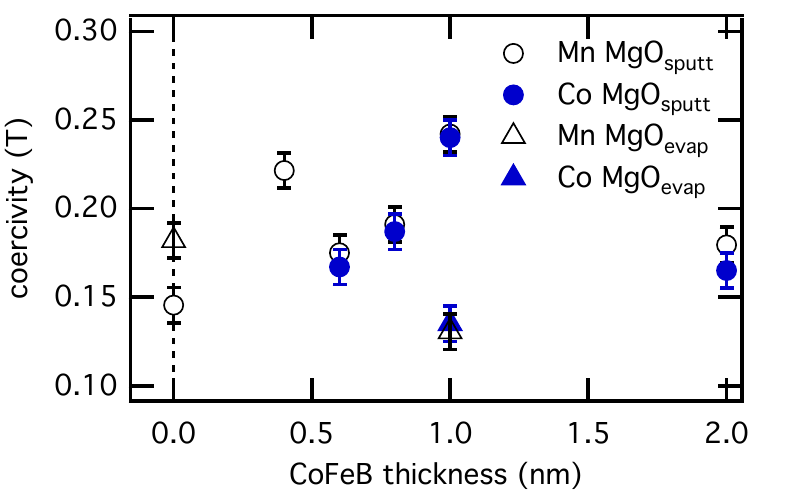}%
\caption{Relation between the \(\rm{Mn}\) and \(\rm{Co}\) coercieve fields for different CoFeB interlayer thicknesses. The orientation of the magnetic field was perpendicular to the film plane. The circles indicate the Mn and Co coercivity for samples with magnetron sputtered MgO and the triangles the samples with e-beam evaporated MgO.}%
\label{fig:Mn_Co_Hc_perp}%
\end{figure}

The film compositions were determined by X-ray fluorescence (XRF) as \(\mathrm{Mn}_{1.63\pm0.06}\mathrm{Ga}\).
\begin{figure}[t!]%
\includegraphics[width=\linewidth]{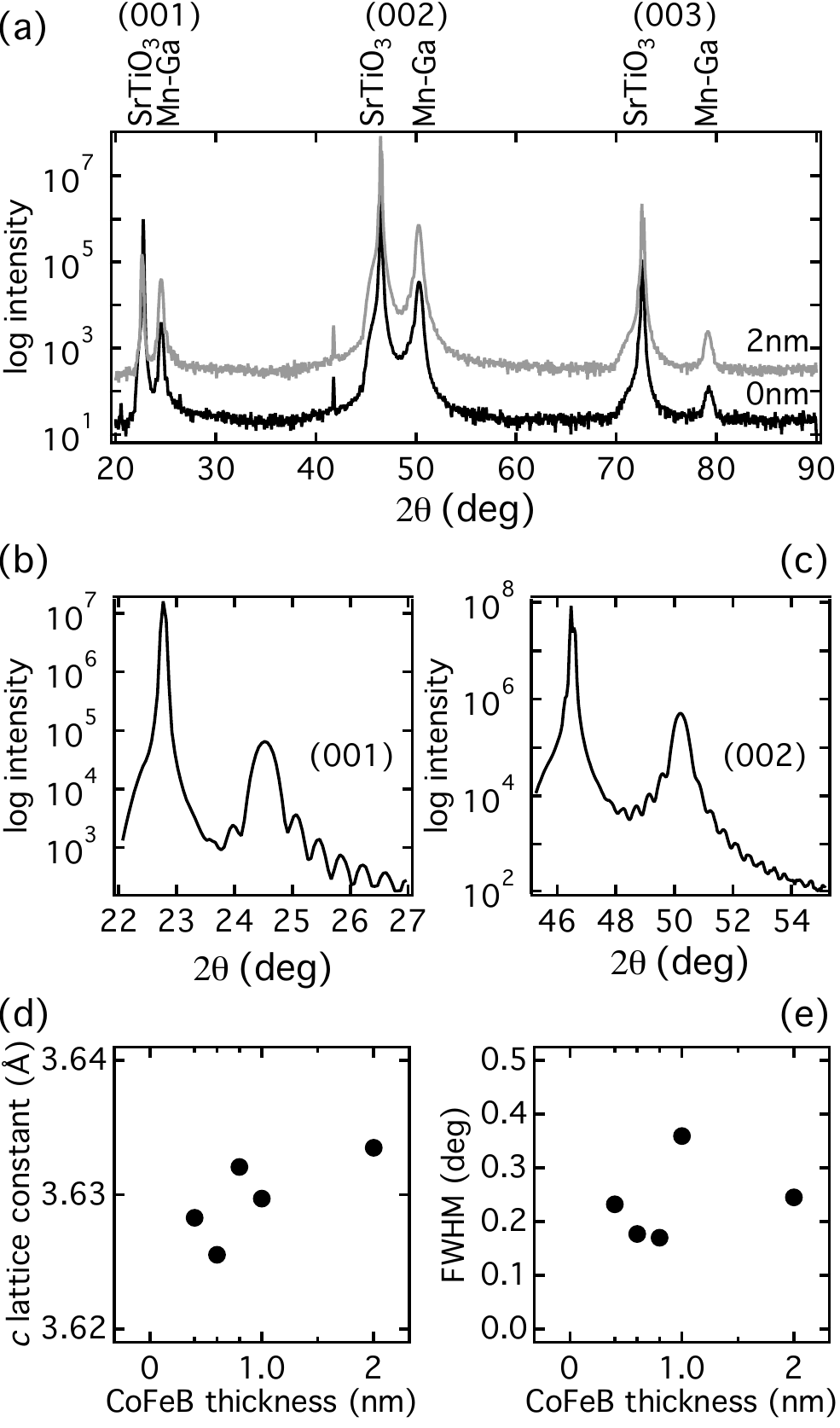}%
\caption{(a) X-ray diffraction scans of Mn\(_{1.63}\)Ga thin films with CoFeB interlayer thickness of 0 and 2\,nm. (b) Laue oscillations on the (001) and (c) on the (002) film peak. (d) \textit{c} lattice constant and (e) full-widths at half-maximum (FWHM) of the rocking curves taken from the (002) reflex.}%
\label{fig:XRD_STO}%
\end{figure}%
XRD measurements (\(\mathrm{Cu\;K\alpha}\) 1.5418\,\AA) were utilized to investigate the crystallographic structure and to determine the crystal phase. Fig.\ \ref{fig:XRD_STO} (a) shows the X-ray diffraction results for the 23\,nm thick Mn\(_{1.63}\)Ga thin films with a CoFeB interlayer thickness of 0 and 2\,nm. All samples show the fundamental (002) peak, corresponding to \(c \approx 3.63\)\,\AA~(see Fig.\ \ref{fig:XRD_STO} (d)). In addition the two superlattice peaks, (001) and (003), are also visible. 
The (011) reflex is forbidden in the L1\(_0\) crystal structure.\cite{Mizukami:2011ft} Pole figure measurements were carried out at a fixed angle of \(2\theta = 33.65\,\rm{deg}\). For each angle \(\Psi\) between 0 and 90\,deg the sample was rotated by \(\Phi = 0,\ldots,360\,\rm{deg}\). The scans showed no appearance of the (011) peak. Laue oscillations, that only appear for thin films with good crystalline coherence, were observed on the superlattice (001) and fundamental (002) peak for all samples. We thus conclude, that our samples show a L1\(_0\) structure with very good crystalline quality. In Fig.\ \ref{fig:XRD_STO} (b) and (c) a fine scan of the (001) and (002) film reflex for a sample with 0.8\,nm CoFeB is shown. At smaller angles the amplitudes of the oscillations are smaller compared to the substrate reflex intensity and therefore not visible. However, due to the weak asymmetry of the fringes we conclude that an almost homogeneous strain along the growth direction is present.\cite{Meinert:2011cy} In addition, rocking curves of the (002) reflex were carried out. The full-widths at half-maximum (FWHM) of rocking curves at the (002) reflex are shown in Fig.\ \ref{fig:XRD_STO} (e). An average value of 0.24\,deg was found showing a narrow distribution of the orientation of the grains. Considering a Gaussian function to describe the peak (shape factor \(K = 0.89\)) a lateral grain size between 25 and 55\,nm was determined.
\newline
\begin{figure}[t!]%
\includegraphics[width=\linewidth]{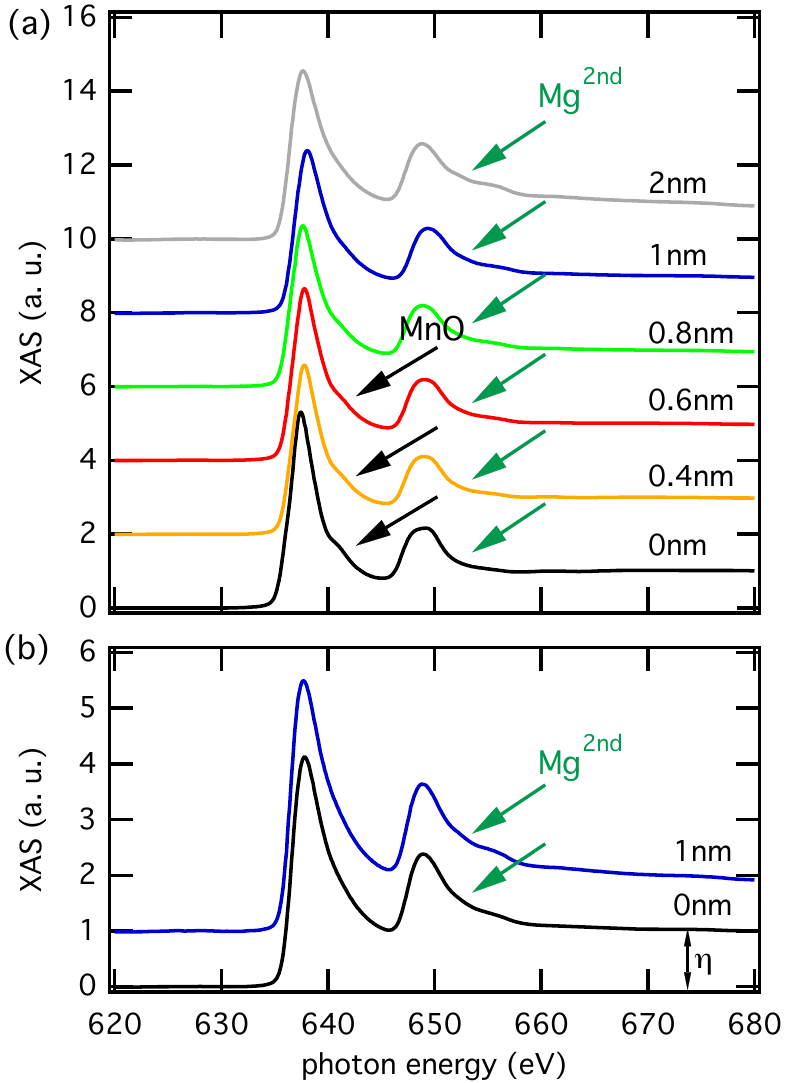}%
\caption{Normalized XAS of Mn on SrTiO\(_3\) substrate. The MgO barrier was deposited by (a) RF magnetron sputtering and (b) e-beam evaporation.}%
\label{fig:XAS_Sput_Evap}%
\end{figure}%
The valence states and magnetic properties of Mn near the Mn-Ga/(CoFeB)/MgO interface were investigated by X-ray absorption spectroscopy (XAS) and X-ray magnetic circular dichroism (XMCD). Fig.\ \ref{fig:XAS_Sput_Evap} (a) depicts the typical absorption spectra of \(\rm{Mn}\) for samples with different \(\rm{CoFeB}\) interlayer thicknesses and RF sputtered MgO. For an interlayer thickness below 0.8\,nm the shape of the \(\rm{Mn}\) spectra slightly differs from those of \(\rm{Mn}\) in the metallic state.\cite{THOLE:1985uc}  At photon energies above the \(L_3\) resonance, an additional feature was found (marked by black arrow) indicating the formation interfacial MnO for CoFeB interlayers thinner than 0.8\,nm. In Fig.\ \ref{fig:XAS_Sput_Evap} (b) the XA spectra for \(\rm{Mn_{1.63}Ga}\) thin films capped with e-beam evaporated MgO are shown with and without a 1\,nm CoFeB interlayer. In this case no evidence for the formation of interfacial MnO was found. Therefore the method used for the deposition of the MgO barrier determines the oxidation of the underlying metal electrode. Furthermore a second-order Mg peak (\(K_1 = 1303.0\, \mathrm{eV}\)) caused by the higher order contributions of the beamline and the MgO capping, was found for all samples above the \(L_2\) edge (marked by green arrow).
\begin{figure}[t!]%
\includegraphics[width=\linewidth]{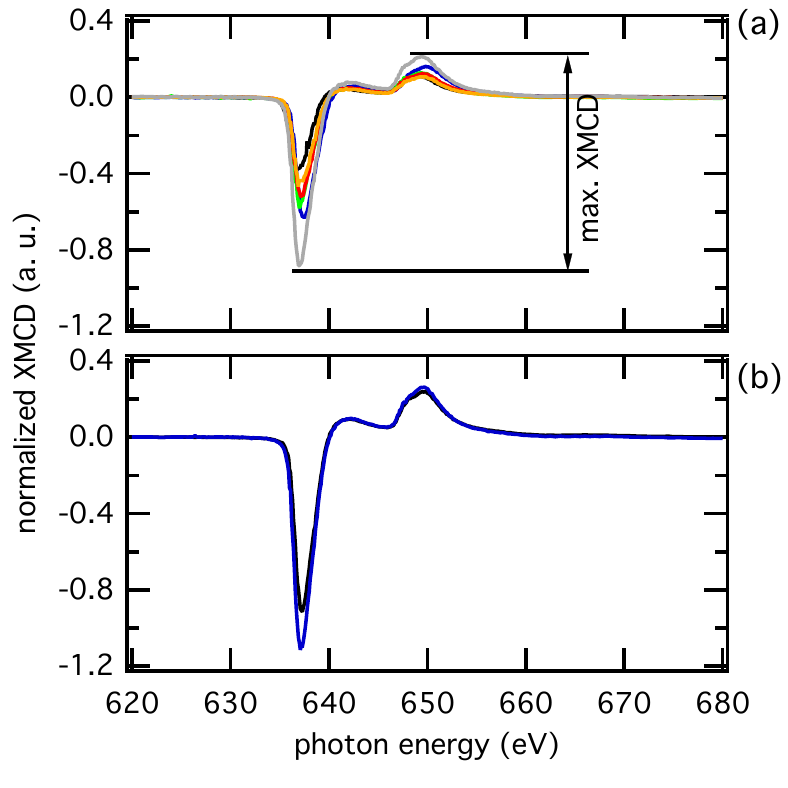}%
\caption{Normalized XMCD spectra (\(\eta = 1\)) of Mn on SrTiO\(_3\) substrate determined from the XAS spectra measured at BL 6.3.1. The MgO capping layer was deposited by (a) RF magnetron sputtering and (b) e-beam evaporation.}%
\label{fig:XMCD_Sput_Evap}%
\end{figure}%
\begin{figure}[h!]%
\includegraphics[width=\linewidth]{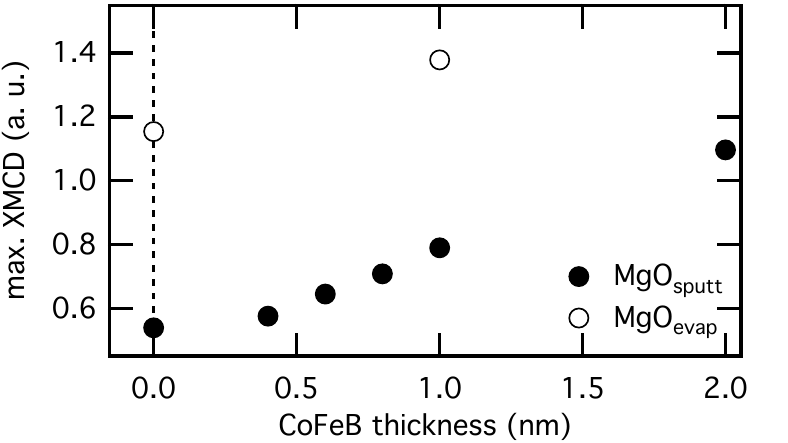}%
\caption{Maximum XMCD signal, as defined in Fig. \ref{fig:XMCD_Sput_Evap} (a), as function of the CoFeB interlayer thickness for thin films with RF magnetron sputtered MgO capping (solid dots) and e-beam evaporated MgO capping (open circles).}%
\label{fig:normXMCD}%
\end{figure}%
Fig.\ \ref{fig:XMCD_Sput_Evap} (a) shows the corresponding XMCD signal normalized to the postedge jump height (i.\ e.,\ \(\eta = 1\)), which increases with increasing CoFeB thickness. Thus, the magnetic moment, which corresponds to the maximum XMCD signal, increases due to the lower amount of MnO. The XMCD data for thin films capped with e-beam evaporated MgO (Fig.\ \ref{fig:XMCD_Sput_Evap} (b)), show the same behaviour as previously seen for \(\rm{Mn_{1.63}Ga}\) thin films with sputtered MgO capping. Therefore the normalized XMCD data agree with the XAS measurements and show, that the Mn magnetic moment at the interface increases with increasing CoFeB thickness.\newline
Fig.\ \ref{fig:normXMCD} shows the maximum XMCD signal, as defined in Fig.\ \ref{fig:XMCD_Sput_Evap} (a), depending on the CoFeB interlayer thickness. The solid dots corresponds to the samples with magnetron sputtered and the circles samples with e-beam evaporated MgO barrier. The XMCD amplitudes and therefore the Mn magnetic moments for magnetron sputtered MgO barrier are lower compared to samples with e-beam evaporated MgO. Even for a CoFeB thickness of 2.0\,nm the XMCD signal is lower for the samples with sputtered MgO. This behaviour gives evidence for a relation between the amount of MnO and the deposition method of the MgO barrier.\newline
The max.\ XMCD signal for the sample without CoFeB interlayer and e-beam evaporated MgO is twice as high as the signal for the sample with magnetron sputtered MgO. Therefore 50\,\% of the signal is suppressed by the MnO. Because the typical probing depth for total electron yield measurements is approximately 1-2\,nm and the resulting thickness of the oxidized Mn layer is approximately 1.4\,nm. Thus, this is an estimate for the maximum amount of MnO in the sample with the maximum amount of MnO.\newline
\begin{figure}[t!]%
\includegraphics[width=\linewidth]{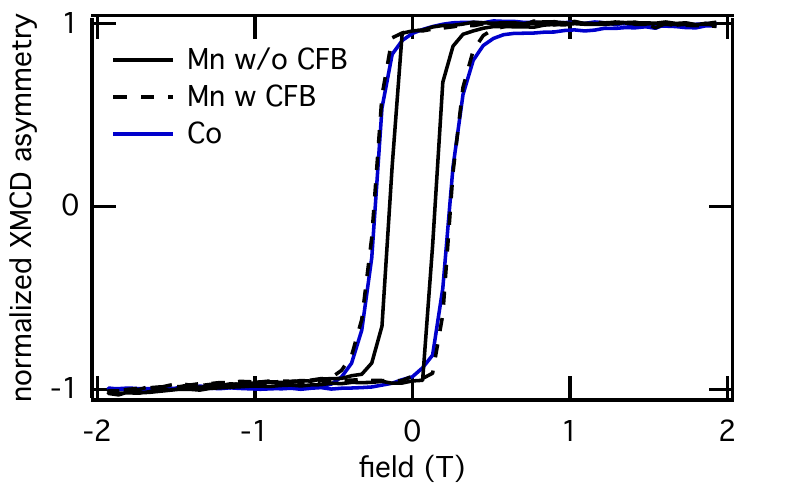}%
\caption{Element-specific hysteresis loops. The dashed black (bold black) curve is the normalized Mn XMCD asymmetry for a sample with (without) 1\,nm \(\rm{CoFeB}\), measured at room temperature for normal incidence of the X-rays and out-of-plane magnetic field. The blue curve is the normalized Co XMCD asymmetry. The MgO protection layer was deposited by magnetron sputtering.}%
\label{fig:Mn_hysteresis}%
\end{figure}%
In Fig.\ \ref{fig:Mn_hysteresis} the element-specific full hysteresis loop of a \(\rm{Mn_{1.63}Ga}\) sample with (dashed black line) and without (bold black line) CoFeB interlayer is shown. In addition the Co hysteresis loop (blue line) is added to compare the coercivity fields of Mn and Co. The \(\rm{Mn}\) hysteresis loop of a sample without interlayer shows a lower coercivity compared to the one with interlayer.
The \(\rm{Co}\) hysteresis loop exhibits the same coercive field as the Mn layer. This is a clear indication that the magnetization of the 1\,nm CoFeB interlayer is perpendicularly aligned to that of the Mn-Ga thin film surface and parallel to the \(\rm{Mn}\) magnetization.
In Fig.\ \ref{fig:Mn_Co_Hc_perp} the coercivity of the Mn and Co films depending on the CoFeB thickness is depicted. The full hysteresis loop of a sample with a 0.4\,nm CoFeB thin film was not detectable for Co due to the small amount of Co in these samples.
For a CoFeB thickness of 0.6 and 2\,nm a Co hysteresis was obtained, but the coercivity slightly differs from that shown by the Mn hysteresis. However, samples with 0.8 and 1.0\,nm CoFeB interlayer thickness showed the same coercivity for Mn and Co. Therefore up to a CoFeB thickness of 2\,nm the interlayer is magnetically aligned with the Mn-Ga bottom layer. The perpendicular magnetization of the CoFeB interlayer originates from three effects: the stray field of the Mn-Ga film, the exchange coupling at the interface, and from the perpendicular magnetic anisotropy of the CoFeB/MgO capping of the Mn-Ga bottom layer. Ikeda et al.\ reported a PMA for CoFeB/MgO/CoFeB tunnel junctions thin film with a CoFeB thickness between 1 and 1.2\,nm.\cite{Ikeda:2010iz} The influence of exchange coupling and the PMA decrease with increasing CoFeB thickness, whereas the stray field is proportional to the film thickness of the underlying \(\rm{Mn}_{1.63}\rm{Ga}\) and remains constant. Thus the range of perpendicular magnetization of the CoFeB is extended to thicknesses up to 2\,nm for our film system.

\section{Conclusion}
We prepared Mn\(_{1.63}\)Ga/CoFeB/MgO thin films and investigated the influence of a CoFeB interlayer on the amount of MnO and therefore on the total magnetic moment at the interface of \(\rm{Mn_{1.63}Ga}\) thin films with CoFeB/MgO. In addition, two different deposition methods for the MgO barrier were examined. Thin films with magnetron sputtered MgO capping and a CoFeB thickness below 0.8\,nm show a difference in the electronic structure of the unoccupied \(3d\) states (see Fig. \ref{fig:XAS_Sput_Evap}) compared to pure Mn which points to MnO formation, whereas for an e-beam evaporated MgO barrier no evidence for MnO was found. The formation of MnO for sputtered MgO can be suppressed by a CoFeB interlayer of at least 0.8\,nm thickness.\newline
In accordance with the XAS, the XMCD data show an increase in the total magnetic Mn moment with increasing CoFeB interlayer thickness. This increase in the magnetic moment can be explained by the decreasing MnO amount. However, the samples with e-beam evaporated MgO barrier exhibit a higher magnetic moment compared to samples with magnetron sputtered MgO barrier. This effect should be taken into account for the further preparation of magnetic tunnel junctions with Mn-Ga electrode.
Element specific full hystereses loops were taken to determine the magnetization axis of the Mn and Co, respectively. For CoFeB thicknesses from 0.8 to 2.0\,nm a perpendicular magnetization of the CoFeB to the sample surface was found. The origin of the out-of-plane magnetization of the CoFeB needs to be investigated further because it is not clear, if the magnetic stray field of the Mn-Ga or an exchange coupling between both materials along with a perpendicular anisotropy introduced by the CoFeB/MgO interface cause the perpendicular magnetization. 

\section{Acknowledgments}
The authors gratefully acknowledge financial support by the Deutsche Forschungsgemeinschaft (DFG, contract number SCHM 1690/6-1 and RE 1052/22-1) and the opportunity to work at BL 6.3.1 of the Advanced Light Source, Berkeley, USA, which is supported by the Director, Office of Science, Office of Basic Energy Sciences, of the U.S. Department of Energy under Contract No. DE-AC02-05CH11231. DE is financially supported by the ERC Advanced Grant (291472 Idea Heusler). 

\bibliography{bibfile}
\bibstyle{paperstyle}

\end{document}